\documentstyle{aipproc}
\begin{document}

\title{GRBs as Hypernovae}

\author{Bohdan Paczy\'nski}
\address{Princeton University Observatory, Princeton, NJ 08544--1001, USA\\
e-mail: bp@astro.princeton.edu}

\maketitle

\begin{abstract}
A standard fireball/afterglow model of a gamma-ray burst relates the
event to a merging neutron star binary, or a neutron star - black hole 
binary, which places the events far away from star forming regions,
and is thought to have an energy of $ \sim 10^{51} $ erg. 
A hypernova, the death of a massive and rapidly spinning star, may
release $ \sim 10^{54} $ erg of kinetic energy by tapping the
rotational energy of a Kerr black hole formed in the core collapse.
Only a small fraction of all energy is in the debris ejected with 
the largest Lorentz factors, those giving rise to the GRB itself,
but all energy is available to power the afterglow for a long time.
In this scenario GRBs should be found in star forming
regions, the optical afterglows may be obscured by dust, and
the early thermal emission of the massive ejecta may give rise to
X-ray precursors, as observed by Ginga.

The optical and X-ray afterglows of GRBs 970228, 97508, 97828 provide
some evidence that these bursts were located in galaxies, most likely
in dwarf galaxies, in or near star forming regions.

\end{abstract}

\section*{Introduction}

The discovery of the
two `gold plated' Gamma Ray Burst (GRB) afterglows: 970228 and 970508,
made a major breakthrough in the field, providing the first direct evidence
for the interaction between the explosive event and its environment, and
in providing the first ever direct distance estimate for one of them,
970508 (Costa et al. 1997, van Paradijs et al. 1997, Sahu et al. 1997, 
Piro et al. 1997, Bond et al. 1997, Metzger et al. 1997ab, Frail et al.
1997, and many presentations at this meeting).
It seems that a now standard fireball/afterglow model agrees
with the observations reasonably well (cf. Wijers et al. 1997,
Waxman 1997, and references therein), though it has a number of explicit
as well as hidden adjustable parameters (cf. Paczy\'nski 1997,
and references therein).  The most popular ultimate energy
source is usually thought to be related to a merging/colliding pair of
neutron stars or a neutron star and a stellar mass black hole, i.e. an
old product of the evolution of a massive binary star.  An alternative
are the `failed SN Ib' (Woosley 1993) and hypernova (Paczy\'nski 1997)
scenarios, in which a young massive and rapidly spinning star,
following the core collapse,
forms a Kerr black hole with a dense torus rotating around it.
Two different modes of energy transfer from the ultimate energy
source (almost always gravitational) to a fireball were proposed:
neutrino - antineutrino annihilations and superstrong magnetic fields.
The kinetic energy injection is considered to be 
$ \sim 10^{51} ~ {\rm erg ~ s^{-1}} $
in a `classical' fireball model, and 
$ \sim 10^{54} ~ {\rm erg ~ s^{-1}} $ in the hypernova scenario.

The afterglow is produced when the ultra-relativistic GRB ejecta
interact with the ambient medium, be it circumstellar, interstellar,
or intergalactic.  To a good approximation the afterglow does not
depend on the nature of the GRB, just on the amount of kinetic
energy released in the explosion.  
In the essential way this is analogous to the formation of
supernovae remnants in the interstellar space, and giant radio lobes
in the intergalactic space, which are the products of supernovae
and galactic nuclear explosions, respectively.

\section*{A Hypernova scenario}

A massive star, at the end of its nuclear evolution, creates an iron core
which is too massive to make a neutron star.  Following the gravitational
collapse a black hole forms.  There are about a dozen binary stars known
with black hole components of $ \sim 10
~ M_{\odot} $ (cf. Tanaka \& Shibazaki 1996, p. 615).
If the star is spinning rapidly then its angular momentum prevents all
matter from going down the drain, and a
rotating, very dense torus forms (Woosley 1993) around the rapidly 
spinning Kerr black hole.  The largest energy reservoir, which may
be accessed with a super-strong magnetic field (cf. Blandford
\& Znajek 1977), is the rotational energy of the black hole:
\begin{equation}
{\rm E_{rot,max} \approx 5 \times 10^{54} ~ 
[erg] ~ \left( { M_{BH} \over 10 ~ M_{\odot} } \right) } .
\end{equation}
The maximum rate of energy extraction by the field was estimated by
Macdonald et al. (1986, eq. 4.50) to be
\begin{equation}
{\rm L_{B,max} \approx 10^{51} ~ [erg ~ s^{-1}] ~ 
\left( { B \over 10^{15} ~ G } \right) ^2
\left( { M_{BH} \over 10 ~ M_{\odot} } \right) ^2 } .
\end{equation}

It is not clear how a superstrong field is generated, even though
it has become popular in theoretical papers over the last few years
(Duncan \& Thompson 1992, Narayan, Paczy\'nski \& Piran 1992, Usov 1992,
Paczy\'nski 1993, Woosley 1993, ....., Woosley 1996, Vietri 1996, 
M\'esz\'aros \& Rees 1997, Paczy\'nski 1997).  The following is a possible
scenario.  A rapidly rotating massive star, just prior to its core collapse,
has a convective shell (Woosley 1997).  According
to Balbus (1997) a large scale magnetic field may be generated in the shell,
and it may reach equipartition with the convective kinetic energy density.
Following the collapse the polar caps of the shell end up in
the black hole, while the equatorial belt becomes part of the torus.
At least two topologies are possible.  In one the
magnetic field lines link the torus to the black hole, while in the other
the field connection is severed.  In both cases
the collapse increases the field strength while the magnetic flux is
conserved, and a substantial radial component can lead to rapid field
increase driven by differential rotation.  In the second case the 
magnetic field assists torus accretion into the black hole by
redistributing its angular momentum, and helps the release of
its gravitational binding energy in the process.  In the first case
a much larger rotational energy of the black hole can be extracted by
the Blandford \& Znajek (1977) mechanism.

It is far from clear if this scenario can work, and if yes, then how
often does it happen?  For the rate to be relevant to GRBs it has to
be $ \sim 10^4 $ times less common than ordinary supernovae.  It may take
a very long time to find the answer with model computations, as two decades
of massive effort have not provided a clear picture of the `core bounce'
that is needed for Type II supernovae.  It may be more productive to 
seek observational evidence for or against this scenario.

A pre-hypernova must be a member of a short period massive binary, so that
tidal interaction can keep the star in a synchronous, i.e. rapid
rotation.  Examples of such systems are some Wolf-Rayet binaries, and in 
particular Cyg X-3, with its $ \sim 5 $ hour orbit.

The death of a massive star cannot be more than a few million years
away from its birth time, and therefore it explodes within its star forming
region, or very close to it.  This makes it distinct from a
popular merging neutron star model, as the merger follows
a long gravitational radiation driven orbital evolution, so it is likely to
be $ \sim 10^9 $ years after the original binary had formed.
During this time the system traveled tens of kiloparsecs, having
acquired a high velocity during the two supernovae explosions (Tutukov
\& Yungelson 1994).

The star forming site for the GRBs in the hypernova scenario implies
that on many occasions the optical afterglow may be heavily obscured
by the dust commonly present in such regions (Jenkins 1997).
Gradual emergence of the fireball out of the
circum-stellar dust shell may affect the early 
afterglow, possibly accounting for the early rise in the
970508 optical light curve.

In a hypernova scenario a small fraction of all kinetic energy
is in the debris ejected with the largest Lorentz factor required
to generate gamma-ray emission, but the bulk
of ejecta is less relativistic, or even sub-relativistic.  Note, that
if $ \sim 10 ~ {\rm M_{\odot}} $ is given $ \sim 10^{54} $ erg then
a typical velocity is $ \sim 10^{10} ~ {\rm cm ~ s^{-1} \approx c/3 } $.
When the fireball is gradually slowed down by the ambient medium the
slower moving ejecta gradually catch up, and they provide a long
lasting energy supply to the afterglow, much larger than the original one
related to the GRB shell.  Therefore, the
afterglow may persist for much longer than predicted by the standard
fireball model.  In case of the nearest GRBs, even thermal
X-ray emission from the hypernova remnant may be detectable, acting
as a calorimeter, and providing a fairly direct measurement of the
total energy release.  

\section*{GRB environment}

The afterglows of three GRBs: 970228, 970508, and 970828, provide some
information about their environment.

{\bf GRB 970228.}  The fading optical afterglow is located near the edge
of a fuzzy $ \sim 25 $ mag object, most likely a dwarf galaxy at a moderate
redshift (Fruchter et al. 1997, and references therein).

{\bf GRB 970508.}  The absorption (Metzger et al. 1997a) and emission
line (Metzger et al. 1997b) redshifts are identical: $ z = 0.835 $, but
the emission line region is not resolved by the HST (Fruchter, Bergeron, \&
Pian 1997).  The compactness of the emission line region makes it very
likely that it is related to the GRB.  Its small size indicates it is
a dwarf galaxy, possibly a star forming region.

{\bf GRB 970828.}  No optical afterglow was detected, and the upper limits are
very stringent (van Paradijs 1997).  The absence of optical emission may be
explained by dust extinction (Jenkins 1997).  Indeed, the X-ray spectrum
indicates the presence of absorbing gas, with $ N_H = 4 \times 10^{21} ~
{\rm cm^{-2}} $, assuming no redshift (Murakami 1997).  The source is at a
high galactic latitude, hence the absorption is likely to be at a large
redshift, possibly near the source.  If the redshift $ z \approx 1 $ is
adopted then the absorbing gas column density is higher by a factor $ \sim 10 $,
and the observed R-band is in the near UV at the source.  Adopting a standard
dust to gas ratio provides more than enough extinction to make the
optical afterglow undetectable.  If the absorbing gas is near the GRB
then this GRB was located in a distant galaxy, possibly in a star forming
region.

While the evidence for the GRBs to be located near star forming regions
is marginal, it will become robust, one way or the other, when a few
dozen of new afterglows are detected.  The answer to the question:
are GRBs related to star forming regions will be answered with future
observations, independently of speculative theoretical models.  


\end{document}